# Measure electron beam energy with the time-resolved beam parameters measurement system and magnetic analyzer mode *

**Wang Yuan**（王远）**Jiang Xiaoguo**（江孝国），**Yang Guojun**（杨国君），**Chen Sifu**（陈思富），
**Xia Liansheng**（夏连胜），**Li Jin**（李劲），**Wei Tao**（魏涛）

Institute of Fluid Physics, China Academy of Engineering Physics, P.O.Box 919-106, Mianyang 621900,China

**Abstract:** Time-resolved beam parameters measurement system can be achieve electron beam energy in LIA. In this paper introduce work principle and method of time-resolved beam parameters measurement system. Show the experiment layout of energy measuring。The principle of magnetic analyzer was described, with the bending radius of 300 mm and the bending angle of 60° after hard-edge approximation, The measured energy is about 18MeV, the maximum energy variation is 2%.

**Key words:** Linear induction accelerator;Time-resolved beam parameters measurement system;Magnctic analyzer; Energy; Energy spectrum

**PACS:** 29.25.Dz, 29.27.Ac, 41.85.Ar

## 1  Introduction

The electron beam energy and energy spectrum is major parameter, it almost is indispensable to study the physical properties and phenomena of something for electron beam.As also it is must for debugging research of newly development linear induction acceleration(LIA). As a result of measuring electron beam energy is not immediacy measure form diode voltage in strong current pulse electron beam LIA, So measure electron beam energy and energy spectrum with magnctic analyzer method.The electron beam of magnctic analyzer was limited by slit on entry and entry positioning of electron beam was sure. The positioning of electron beam and image width were sure base on time-resolved beam parameters measurement system. The electron beam energy was computed form magnetic field intensity: $\mathbf{E} = 18.2\text{MeV}$. The electron beam energy spectrum was obtained form breadth of image: $\dfrac{\Delta E}{E} \approx 2\%$ [1].

Time-resolved beam parameters measurement and diagnose system was based experiment of electron beam energy measured in newly debugging LIA [2]. The maximum time-resolved capability is 2ns, the sampling ratio of surface is 12～20 point/mm, The system can be wide dynamic range for capture data and obtain data of high dynamic range. So it is Supported for small signal of electron beam was obtained; Develop fiber synchronization trigger system and synchronization monitor system for multi-frame gated open base on pulse power switch system, It supplies credibility guarantee to real time estimate time of electron beam was obtained, synchronization precisioncan be arrive at 1ns base on a mass of experiment data, it is importance meaning for study accurate burst trigger mode of pulse electron beam; The agility multi-frame gated control mode can be ensure to record split second data of all electron beam on any time[3-4].

## 2  Time-resolved beam parameters measurement system

Fig.1 shows the experiment layout of the time-resolved beam parameters measurement system. While current（or beforehand pulse current）is transported from load of pulse power system, current obtain synchronization signal pass by accelerator cell, the signal by way of entry of synchronization trigger system, the signal engendered after some delay to trigger measurement system, the multi-frame gated after inherence delay, if the delay time is right that measurement system can be obtain image of pulse electron beam while experiment result is just on time of the multi-frame camera gated.

Fig.2 shows the frame of optics of time-resolved beam parameters measurement system, the multi-frame camera is part and parcel in this system, it decided capability of time-resolved and image data dynamic range、distinguish rate of space. The multi-frame camera need two input trigger signal at work: External trigger synchronization signal of system(EXT) and image increase gating trigger signal(GATE). ICCD is key part in the multi-frame camera, because of technics and development of optics component, so it make up of microchannel plank(MCP)、coupling lens、CCD camera. Because this camera need higher shutter speed, so image gating component was selected whit high speed MCP in reference[2], its availability cathode diameter is 25mm and best fast shutter time is 2ns, this is use key component in the multi-frame camera[5-8].

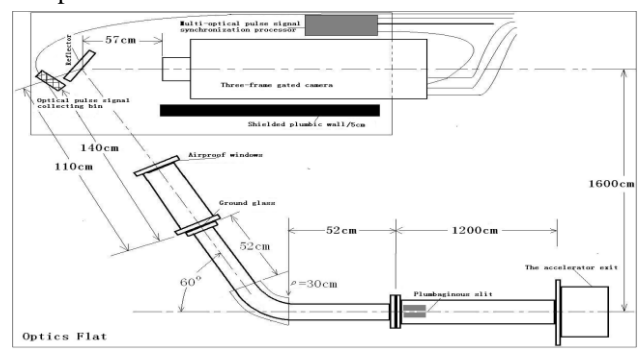

Fig.1 Energy measurement based on magnctic analyzer method

Received   ?
Supported by National Natural Science Foundation of China（10675104，51077119，11375162）
Email: ideawy@163.com.    Tel:0816-2484140

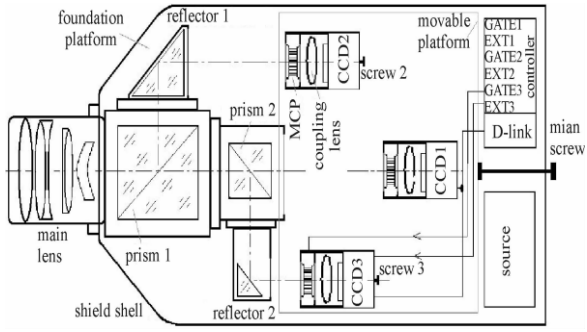

Fig.2 Structrue of framing camera

Table 1 show main performance

Table 1 Main performance of three-frame gated ICCD measurement system

| | Exposure time /ns | interval time/ns | spatial resolution (lp.mm) | equivalent background lamination(EBI)/ (electrons pixel s) | response uniformity /% | linearity /% | inherent insertion delay/ns | image size (in diameter)/mm |
|---|---|---|---|---|---|---|---|---|
| Min: | 2 | ≈0 | | | | | 36.8 | |
| Max: | ~$10^9$ | ~$10^9$ | >30 | 10 | 2 | 1 | 37.3 | 25 |
| tep: | <0.5 | <0.5 | | | | | | |

## 3  Measuring electron beam energy

The electrification particle can be turn in magnetic field, while electron beam into equality magnetic field, beam centric trajectory is circle on magnetic force line vertical plane, as shown in Fig. 3. If the total magnetic field passing through a surface is B(magnetic flux), then electron of standard energy moved track curve of circle. Its tangent directional speed is $v$, if the bending radius is $\rho$, while electron turned along a curving trajectory. Therefore, the centrifugal of $f_e$ can be expressed as[9]:

$$f_e = \gamma m_0 \frac{v^2}{\rho} \quad (1)$$

On the other hand, the electron was produced centripetal force in magnetic field B, this centripetal force of $f_c$ can be expressed as:

$$f_c = -\frac{1}{c} evB \quad (2)$$

Just as balancing $f_e$ and $f_c$, then the electron was controlled to move on track curve of circle, this bending radius is $\rho$, the $B_\rho$ can be expressed as from a mathematical equation $f_e + f_c = 0$:

$$B_\rho = \frac{m_0 c^2 \gamma \beta}{e} \quad (3)$$

Where $B_\rho$ as a rule is magnetic flux, $m_0$ is electron weight, $\gamma$ is gene of theory of relativity, $\beta$ is electron speed, further more, $\beta = \frac{1}{v}g\sqrt{v^2-1}$, $e$ is quantity of electron.

Using Eqs.(3), it can be obtain momentum of electron $P_0$ merely with measure magnetic flux B and bending radius $\rho$, Eqs.(3) has been altered energy equation with this Eqs $v = 1 + E/m_0 c^2$ for know the electron energy E, considering theory of relativity, then E can be expressed as:

$$E = m_0 c^2 [\sqrt{1+(\frac{eB\rho}{m_0 c^2})^2} - 1] \quad (4)$$

Bending radius $\rho$ is fixed in the configuration of magnetic analyzer, then if only need measure magnetic flux B can be achieve electron energy E.

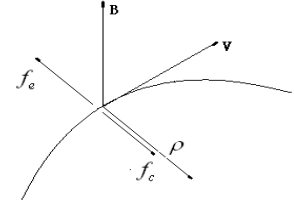

Fig.3 The plan of electron Particle is pressed in magnetic field

The layout of electron beam energy measure as shown in Fig. 4, the magnetic analyzer with the bending radius of 300mm and the bending angle of 60° after hard-edge approximation was installed end of excursion vacuum tube of accelerator, the magnetic analyzer connectted exit of accelerator using excursion vacuum tube of 1.2m long, the slit by 30mm wide located entry of beam after hard-edge approximation of magnetic pole, the ground glass target located exit of excursion vacuum tube after hard-edge of magnetic pole by 52cm, the incident ray and reflection ray of electron beam is all vertical magnetic field of the analyzer. The image was created by the electron beam bombardment target that was turn in magnetic field, the image was obtained base on time-resolved beam parameters measurement system to sure position of track of electron beam. Where the loop of magnetic analyzer supplied stabilization power by constant current, magnetic field intensity was measure with fluxmeter, using measure B and bending radius $\rho$, it can be found from Eq. (4) that compute electron energy is $E \approx 18.2 MeV$.

## 4  Measuring electron beam spectrum

While different momentum of electron passed equality magnetic field, the different electron would move on corresponding the path of bending radius, it obtained corresponding energy spectrum of an electron on exit of magnetic field, this is so-called spectrum,

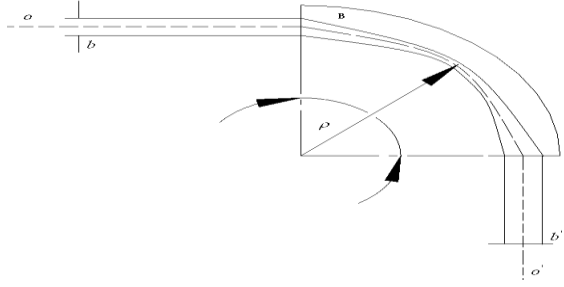

Fig.4 The layout of electron beam energy measure

as a matter of fact is different imaging spot position that is finally the electron beam bombardment targe. Therefore, compare the single electron beam with the electron beam of energy spread, this latter beam can be beam spot image broadened, all appearance electron beam spot image width just rested with[10]:

1、The slit width is b, namely the width of electron beam in entry, Fig.2 shows 'ob' is the half of slit width, if the magnetic field is ensure, while the speed of electron $v_0$ along turn on the path of bending radius $\rho$, the image width $b'$ on exit of magnetic analyzer can be expressed as:

$$b' = Mb \quad (5)$$

Here M is ensure zoom in ratio by characteristic of the magnetic analyzer:

$$M = \frac{b'}{b} = -\frac{f}{l-g} \quad (6)$$

Where $f$ is focus of the magnetic analyzer, $l$ is distance of the object after hard-edge approximation of magnetic pole, g is distance of the focus after hard-edge approximation of magnetic pole,

2、The decentralization of momentum of electron can be expressed as $\Delta v = (v - v_0)/v_0$, here, if the magnetic field is ensure, the electron of decentralization of relativity momentum $\Delta v$ passed the magnetic analyzer, then the image width $b'$ can be expressed as:

$$b' = \rho \Delta v (1-M) \quad (7)$$

In evidence, usually after the beam of electron passed the magnetic analyzer, the all image width is the sum of hereinbefore two parts.

$$b = Mb + \rho \Delta v (1-M) \quad (8)$$

According to relation of momentum and energy know that energy spectrum is $\varepsilon = \Delta E/E = 2\Delta P/P = 2\Delta v$, so Eqs.(8) can be rewrite this expressed as:

$$b' = Mb + \frac{1}{2}\varepsilon\rho(1-M) \quad (9)$$

But for compute and measure convenience, usually using all slit to expresse the energy spectrum, using $W = 2b, W' = 2b'$ to exchange Eqs.(9), then the energy spectrum can be expressed as:

$$\frac{\Delta E}{E} = \frac{4W'}{\rho(1-M)} \quad (10)$$

In this way is need measure image width $W'$ can be compute corresponding energy spectrum. Fig.5 shows the time-resolved beam parameters measurement system captured image spot of beam spectrum on the ground glass target, images of pulse electron beam captured with 10ns shutter in three measure flat moment of electron beam (the former、the middle、the end), corresponding delay time of the measure system are 130、190、210ns, measured corresponding image width are 1.96、1.89、1.78mm, time after time measure indicated that the energy spectrum of electron beam is $\frac{\Delta E}{E} \leq 2\%$.

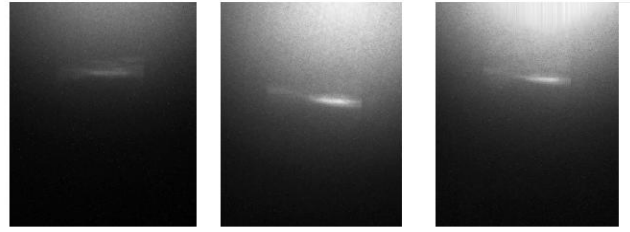

Fig.5 Images of slit beam captured with 10ns shutter

## 6 Discussion

Although the magnetic field of magnetic analyzer was builded base on exact machining magnetic pole, but the existence of hard-edge of magnetic field is not hard certain, the magnetic leaked is difficult computed, and that magnetic field B and bending radius $\rho$ are all difficult accurately ensure, the magnetic field of path be likely to partial asymmetry. In addition, the relation between energy E and magnetic field B is nonlinear, the exactitude of magnetic analyzer adjust is not easy. So the magnetic analyzer usually is not use for energy of electrification particle absoluteness measure. In our formerly experiment, although the energy of pulse electron beam measure was not very exactitude, but measure exactitude was advanced by using time-resolved beam parameters measurement system. The measuring errors can in this way compute, if magnetic field current was very stabilization、magnetic pole machining exactitude、magnetic field of path thinked symmetrical 、value of B looked upon stabilization and symmetrical, measuring errors of magnetic field edge effect is ~5%, measuring errors of path is ~3.5%, then the total measuring errors of electron energy is less than 10%. Moreover add reading errors of measuring energy spectrum is~3.5%, the total measuring errors of electron energy is less than 12%.